\documentclass{llncs}
\usepackage{makeidx}  
\usepackage{amsmath}
\usepackage{amsfonts,amssymb}
\usepackage{clrscode3e}        
\usepackage{multirow}
    
\newcommand{\dt}{\displaystyle}

\begin{document}
\frontmatter          
\pagestyle{headings}  
\addtocmark{The set cover problem} 
\title{An estimation of the greedy algorithm's accuracy
for a set cover problem instance}
\titlerunning{Accuracy estimating of the greedy solution}  
%
\author{Alexander Prolubnikov}
\authorrunning{Alexander Prolubnikov} 
%
\tocauthor{Alexander Prolubnikov}
\institute{Omsk State University, Omsk, Russian Federation\\
\email{a.v.prolubnikov@mail.ru}
}
\maketitle              

\begin{abstract}

Considering the set cover problem, by modifying the approach that gives a logarithmic approximation guarantee for the greedy algorithm, we obtain an estimation of the greedy algorithm's accuracy for a particular input. We compare the presented estimation to another estimations of this type. We give such examples of the set cover problem instances that the presented estimation sagnificantly 
improves over linear programming relaxation based estimation.

\end{abstract}

\section{The set cover problem}

In the set cover problem ({\it\!SCP}), we have a set
$U\!=\!\{1,\ldots,m\}\stackrel{\text{\tiny dn}}{=}[m]$, and 
such a collection of its subsets $S\!=\!\{S_1,\ldots, S_n\}$ that 
$$\bigcup\limits_{i=1}^n S_i=U.$$ The collection of sets
$S'\!=\!\{S_{i_1},\ldots, S_{i_l}\}$, $S_{i_j}\!\in\!S$, is called 
a {\it cover} of $U$ if $$\bigcup\limits_{j=1}^{l}S_{i_j}=U.$$ 
We have a {\it weight function} $w\!:\!S\!\rightarrow\!\mathbb{R}_+$ 
($\mathbb{R}_+\!=\!\{x\!\in\! \mathbb{R}:x\!\ge\! 0\}$), $w_i\!\stackrel{\text{\tiny dn}}{=}\!w(S_i)$ is
a {\it weight} of the set $S_i$. The weight of the collection of sets 
$S'\!=\!\{S_{i_1},\ldots, S_{i_l}\}$
is equal to the sum of weights of the sets that it contains:
$$w(S')\!=\!\sum\limits_{j=1}^l w(S_{i_j}).$$ To solve the problem, 
we must find an optimal cover of $U$, i.e. we need to find the cover 
of $U$ that has minimum weight. 

{\it An instance} is an SCP with predefined $U$, $S$ 
and $w$. Let $\mathcal{A}$ be an approximate algorithm for SCP. {\it An approximation 
guarantee} of $\mathcal{A}$ is the value $\rho_{\mathcal{A}}(m)$ such that, for every 
instance $\mathcal{P}$ that may be defined on $U\!=\![m]$, we have 
$$\frac{w(Cvr)}{w(Opt)}\le\rho_{\mathcal{A}}(m),$$ where $Cvr$ is a cover obtained 
by $\mathcal{A}$ for the instance $\mathcal{P}$ and $Opt$ is an 
optimal solution of $\mathcal{P}$. 

SCP is $NP$-hard \cite{Garey}. In \cite{Dinur}, it has been shown that,
whenever $P\!\neq\!NP$ holds, $\rho_{\mathcal{A}}(m)\!>\!(1-o(1))\ln m$ for any
approximate algorithm $\mathcal{A}$ with polynomial complexity. In \cite{RazSafra,Feige}, 
there have been presented another inapproximability results for SCP which 
exclude the possibility of a polynomial time approximation with better than 
logarithmic approximation guarantee. 

There is the greedy algorithm for approximation of set covering
with complexity $O(m^2n)$. For positive real valued weights, it holds 
\cite{Chvatal} that $$\dt\frac{w(Gr)}{w(Opt)}\le H(m)\le \ln m+1,\eqno(1)$$ 
where $Gr$ is the cover that the algorithm produces 
and $H(m)\!\stackrel{\text{\tiny dn}}{=}\!\sum_{k=1}^m1/k$. For a particular 
instance of the problem, we have
$$\dt\frac{w(Gr)}{w(Opt)}\le H(\tilde{m}),\eqno(2)$$ 
where $\tilde{m}\!\stackrel{\text{\tiny dn}}{=}\!\max\{|S_i|:S_i\!\in\!Opt\}$ 
and it is specified by the instance. 
Since SCP is $NP$-hard, instead of $\tilde{m}$, we must use the value of 
$\bar{m}\stackrel{\text{\tiny dn}}{=}\max\{|S_i|:S_i\!\in\!S\}$ in order 
to obtain more tight upper bound 
on the ratio $w(Gr)/w(Opt)$ for a given instance. For non-weighted case of SCP, the upper and lower 
bounds on the approximation guarantee have been obtained in \cite{Slavik}. 
It has been shown that, for the worst case $\mathcal{P}$ that may be specified 
on $U\!=\![m]$, we have
$$T_l(m)<\dt\frac{w(Gr)}{w(Opt)}<T_u(m),$$
where $T_l(m)=\ln m-\ln\ln m-0.31$, $T_u(m)=\ln m-\ln\ln m+0.78$.

\bigskip

Considering a particular instance of the problem, we show how to estimate 
the ratio $w(Gr)/w(Opt)$ more precisely than the common logarithmic 
approximation guarantee (1) suggests. The presented estimation is more accurate than  
$H(\bar{m})$ for majority of instances for which the value of $\bar{m}$ 
is large enough regarding to $m$. We give such examples of instances that the presented estimation sagnificantly improves over linear programming relaxation based estimation. The estimation we present may be used to obtain the lower bound on the 
optimal cover weight and so it can be applied to branch and bound strategies
for the problem.

\section{The estimation of the greedy algorithm's accuracy for an instance of the problem}

Implementing the greedy algorithm, we take the sets from $S$ 
into $Gr$ relying on the values of {\it charged weights} $w_i/|S_i|$, 
i.e. using the values of the weights that $S_i$ charges 
to yet uncovered elements of $U$ at the moment we choose a set to include
into $Gr$.

\begin{center}
\fbox{\parbox{\textwidth}{
\begin{center}
{\bf The greedy algorithm for SCP}\\
\end{center}

\quad {\bf Step 0.} $Gr\!:=\!\varnothing$.

\quad {\bf Step 1.} If $Gr$ is a cover of $U$, then stop the algorithm, 
else go to the Step 2.

\quad {\bf Step 2.} Choose such $k$ that
$$\dt\frac{w_k}{|S_k|}=\min\biggl\{\frac{w_i}{|S_i|}\ :\	\biggl (S_i\!\in\! S \biggr ) \wedge
\biggl (S_i\!\not\subset\!\bigcup\limits_{S_j\in Gr}S_j\biggr )\biggr\},\eqno(3)$$
\qquad\qquad\quad $Gr\!:=\!Gr\!\cup\!\{S_k\}$. $S_i\!:=\!S_i\!\setminus\! S_k$
for all $S_i\!\not\in\!Gr$. Go to the Step 1.\\

}}

\end{center}

Let it takes $l$ iterations of the greedy algorithm to cover $U\!=\![m]$ for 
the instance $\mathcal{P}$ of SCP. Let the algorithm covers $s_k$ elements of $U$ 
on its $k$-th iteration and let $m_k$ denotes the number of yet uncovered elements 
of $U$ after the $k$-th iteration is completed, $m_0\!=\!m$, and let 
$s\!=\!\{s_1,\ldots,s_l\}$. We shall prove the following theorem.

\bigskip

\noindent {\bf Theorem.} $$\dt\frac {w(Gr)}{w(Opt)}\le G(s)=H(m)-\Delta(\mathcal{P}),\eqno(4)$$ 
{\it where $$G(s)\stackrel{\text{\tiny dn}}{=}\sum\limits_{k=1}^l\frac{s_k}{m_{k-1}},$$ $\Delta(\mathcal{P})\!\ge\! 0$. 
$\Delta(\mathcal{P})\!=\!0$ if and only if it takes $m$
iterations of the greedy algorithm to obtain a cover of $U$.}

\bigskip

Also we shall write $G(\mathcal{P})$ meaning the same thing as when we write $G(s)$ since the instance $\mathcal{P}$ uniquely identifies the sequence $s$. The estimation $G(s)$ is a refinement of the estimation (1). We prove (4)
by modifying the well known proof of (1). For example, it is presented in \cite{Vazirani}. 
But, estimating the ratio $w(Gr)/w(Opt)$, instead of majorization of the weights 
that sets from $Gr$ charge to distinct elements of $U$, we majorize weights 
of the sets itself. Doing this, we obtain the estimation (4) for an instance and this estimation appears to be more accurate than $H(\bar{m})$ for 
a wide share of instances.

\smallskip

Suppose that, before implementing the $k$-th iterartion of the greedy 
algorithm, we have the subset $U_k$ of the elements of $U$ that are 
yet uncovered, $m_k\!=\!|U_k|$. After implementing the Step 2
on the previous iterations, all of the sets in $S$ contain only
elements of $U_k$. Let us suppose that, in accordance with (3), the greedy 
algorithm chooses the set $S_k$ at the $k$-th iteration. Then, as it will be
shown further,  
$$\dt\frac{w(S_k)}{|S_k|}\leq\dt\frac{w(Opt_k)}{m_k},$$ 
where $Opt_k$ is an optimal cover of $U_k$ that we may obtain using
the modified sets from $S$. We use this inequality to prove (4). It is also 
used to prove (1) and (2), but we prove the different 
statement that deals with another subject. We don't search for the worst 
case instance of the problem to obtain an estimation of the greedy algorithm's accuracy on it,
but, for a particular instance of the problem, we estimate accuracy of the greedy algorithm. And, as a result, we obtain a bound of a different type on $w(Gr)/w(Opt)$.

\bigskip

For an instance $\mathcal{P}$ with $U\!=\![m]$, let the ordered 
collection of sets $\{S_1,S_2,\ldots,S_l\}$ be the cover $Gr$. 
Let $Opt\!=\!\{A_1,A_2,\ldots,A_r\}$, $A_i\!\in\!S$, be an optimal 
cover and let $a_i\!\stackrel{\text{\tiny dn}}{=}\!w(A_i)$.

\bigskip

\noindent {\bf Lemma.} 
$$\dt\frac {w(S_1)}{|S_1|}\leq\dt\frac {w(Opt)}{m}.$$

\bigskip

\noindent {\bf Proof.} 
Let $$A_1'\!=\!A_1,\ A_2'\!=\!A_2\!\setminus\! A_1,\ \ldots,\
A_r'\!=\!A_r\!\setminus\bigcup\limits_{j=1}^{r-1}A_j.$$ The sets
$A_j'$ are pairwise disjoint. Let us renumber the sets $A_j'$ 
in accordance with nondecreasing order of ratios $a_j/ |A_j'|$. 
Thus we have $$\dt\frac{a_1}{|A_1'|}\leq\dt\frac{a_2}{|A_2'|}\leq\ldots\leq
\dt\frac {a_r}{|A_r'|}.$$ Since, for positive $a,b,c,d$, holds
$$\biggl (\dt\frac{a}{c}\leq\dt\frac{b}d\biggr )\ \Rightarrow\ 
\biggl (\dt\frac{a}{c}\leq\dt\frac{a+b}{c+d}\biggr ),$$ then 
$$\dt\frac{a_1}{|A_1'|}
\leq\dt\frac {\sum\limits_{j=1}^r a_j}{\sum\limits_{j=1}^r |A_j'|}=
\dt\frac{w(Opt)}{m}.$$ Taking into account (3), we have
$$\dt\frac{w(S_1)}{|S_1|}\leq\dt\frac{a_1}{|A_1|}=\dt\frac{a_1}{|A_1'|}
\leq\dt\frac{w(Opt)}{m}.$$ 
$\blacksquare$

\bigskip

Let us prove the Theorem now.

\bigskip

\noindent {\bf Proof.} Let us consider a selection of a set in $Gr$ 
at the $k$-th iteration as a selection of the first set to cover the 
part of $U$ that is yet uncovered before the iteration. That 
is to say that we consider a new instance of SCP with modified 
$S$. The weights of the sets are the same as initially. Let the greedy 
algorithm chooses the set $\widehat{S}_k$ on the $k$-th iteration of its 
operating, where $$\widehat{S}_k=S_k\setminus\bigcup\limits_{j=1}^{k-1} S_j.$$ 
Since $s_k\!=\!|\widehat{S}_k|$, using the proven Lemma, for every iteration, 
we have $$w_1\leq \dt\frac{s_1}{m_0}\,w(Opt),\ 
w_2\leq \dt\frac{s_2}{m_1}\,w(Opt_2),\ \ldots,w_l
\leq \dt\frac{s_l}{m_{l-1}}\,w(Opt_l).$$ 
Since $w(Opt_l)\leq\ldots\leq w(Opt_2)\leq w(Opt)$, it holds that
$$w(Gr)=\sum\limits_{k=1}^l w_k\leq \dt\frac{s_1}{m_0}\,w(Opt)+
\dt\frac {s_2}{m_1}\,w(Opt_2)+\ldots+
\dt\frac{s_l}{m_{l-1}}\,w(Opt_l)\leq$$ $$\leq\biggl(\dt\frac{s_1}{m_0}+
\dt\frac {s_2}{m_1}+\ldots+\dt\frac{s_l}{m_{l-1}}\biggr ) w(Opt)=$$ 
$$=\biggl(H(m)-H(m)+\dt\frac {s_1}{m_0}+\dt\frac {s_2}{m_{1}}+
\ldots+\dt\frac {s_l}{m_{l-1}}\biggr ) 
w(Opt)=(H(m)-\Delta(\mathcal{P})) w(Opt),$$ where $$\Delta(\mathcal{P})\!=\!H(m)-\biggl(\dt\frac {s_1}{m_0}+\dt\frac {s_2}{m_{1}}+
\ldots+\dt\frac {s_l}{m_{l-1}}\biggr).$$ 
Assuming $m_l\!=\!0$ and summing over $i$ in deacreasing order, we obtain
$$\Delta(\mathcal{P})=\sum\limits_{k=1}^{l}\biggl (\sum\limits_{i=m_{k-1}}^{m_{k}+1}
\frac{1}{i}-\frac{s_{k}}{m_{k-1}}\biggr ).$$
Since $$\Delta(\mathcal{P})=\sum\limits_{k=1}^{l}\biggl (\sum\limits_{i=m_{k-1}}^{m_{k}+1}\frac{1}{i}
-\frac{s_{k}}{m_{k-1}}\biggr )=\sum\limits_{k=1}^{l}\biggl (\sum\limits_{i=m_{k-1}}^{m_{k}+1}\frac{1}{i}
-\sum\limits_{i=m_{k-1}}^{m_{k}+1}\frac{1}{m_{k-1}}\biggr )=$$
$$=\sum\limits_{k=1}^{l}\sum\limits_{i=m_{k-1}}^{m_{k}+1}\biggl (\frac{1}{i}-\frac{1}{m_{k-1}}\biggr ),$$ 
we have $\Delta(\mathcal{P})\!\ge\!0$ and $\Delta(\mathcal{P})\!=\!0$ 
if and only if $m_{k-1}=m_k+1$, i.e. if $s_k=1$ for all of 
$k\!=\!\overline{1,l}$. $\blacksquare$

\section{Estimating of the greedy algorithm's accuracy\\ using $G(s)$ and other estimations}

Let $\mathcal{P}$ be an instance with $U\!=\![m]$, $S\!\subseteq\!2^U$. 
By the proven theorem, for the instance $\mathcal{P}$, we may estimate the ratio 
$w(Gr)/w(Opt)$ using $G(s)$, where $s$ is a sequence that 
the greedy algorithm produces on $\mathcal{P}$. Let $m(s)\!\stackrel{\text{\tiny dn}}{=}\!\max\{|s_i|:s_i\in s\}$.

\subsection{A comparison of $G(s)$ with $H(\bar{m})$} Let us compare $G(s)$ to $H(\bar{m})$ (but not to $H(m)$) considering them as two different ways of obtaining the bound on $w(Gr)/w(Opt)$ for a particular instance. 
Any bound of such type ought to be compared with $H(\bar{m})$ since, at best, all of them are logarithmic for the worst case instances.

We compare $G(s)$ to $H(\bar{m})$ 
for all of the the instances that may be defined on $U\!=\![m]$ for $m\!=\!\overline{10,35}$, using the classes $\mathcal{C}_s$ of instances. 

\paragraph{Classes $\mathcal{C}_s$ of instances.} For every sequence $s\!=\!\{s_1,\ldots,s_l\}$, there exists such a class 
$\mathcal{C}_s$ of instances that the greedy algorithm 
produces $s$ on them. For example, the greedy
algorithm produces a particular sequence $s$ if the instance has the 
following form. Let $l\!>\!1$, $q_i\!=\!\sum_{j=1}^i\!s_j$, $i\!=\!\overline{2,l}$, $q_1\!=\!0$. 
Let $$S=\{S_1,\ldots,S_l,A\},$$ where $S_i\!=\!\{q_i+1,\ldots,q_i+s_i\}$, 
$A\!=\!\{1,2,\ldots,m\}$. Let $w(S_i)\!=\!|S_i|\!=\!s_i$ for $i\!=\!\overline{1,l\!-\!1}$, 
$w(S_l)\!=\!|S_l|\!+\!1$, $w(A)\!=\!m\!+\!\varepsilon$, $0\!<\!\varepsilon\!<\!1$.
Using the greedy algorithm, we obtain the cover which consists of 
sets $S_1,\ldots,S_l$ for such instances. The weight of this cover is equal to $m\!+\!1$, 
while the weight of the optimal cover (the weight of the set $A$) is equal to 
$m\!+\!\varepsilon$. So, having the instance as an input, the greedy algorithm produces  
the sequence $s$. 

To compare $G(s)$ and $H(\bar{m})$, for any $m$, we split the set of instances on $U\!=\![m]$ into the classes $\mathcal{C}_s$. Since $m(s)\!\le\!\bar{m}$ for any $\mathcal{P}\!\in\!\mathcal{C}_s$, the 
value of $H(m(s))$ is not larger 
than the value of $H(\bar{m})$. Thus, counting the number of the 
instances of $s$ for which $G(s)\!<\!H(m(s))$ holds, 
we obtain a numeric lower bound on the number of such instances $s$ that
$G(s)\!<\!H(\bar{m})$ holds for all $\mathcal{P}\!\in\!\mathcal{C}_s$. 
And thus, for given $m$, we estimate the share of such classes 
$\mathcal{C}_s$ that $G(s)\!<\!H(\bar{m})$ for 
$\mathcal{P}\!\in\!\mathcal{C}_s$.  

\paragraph{The share of classes $\mathcal{C}_s$ for which $G(s)\!<\!H(\bar{m})$.} Let $\mu(s)\!\stackrel{\text{\tiny dn}}{=}\!m(s)/m$.
 In the Table 1, according to $\mu(s)$ that belongs 
to predefined intervals, we show the shares of the classes 
$\mathcal{C}_s$ for which the estimation $G(s)$ appears to 
be more accurate than the estimation $H(m(s))$. 
Going through all of the possible instances of $s$ for $m\!=\!\overline{10,35}$, 
we have found 
that the share of such classes grows as the value of $\mu(s)$ grows.
And it also shows that such a share tends to grow as $m$ grows.

\begin{table}[h!]
\begin{center}
\begin{tabular}{|c||c|c|c|c|c|}
\hline
$m$ & $\mu\!\in\!(0,0.2]$ & $\mu\!\in\!(0.2,0.4]$ & $\mu\!\in\!(0.4,0.6]$ & $\mu\!\in\!(0.6,0.8]$ & $\mu\!\in\!(0.8,1]$\\	
\hline
\hline
10 & 0 & 13.5 & 64.8 & 100 & 100\\
\hline
11 & 0 & 12.4 & 54.7 & 92.5 & 100\\
\hline
12 & 0 & 9.9 & 52.9 & 97.5 & 100\\
\hline
13 & 0 & 13.9 & 64.5 & 93.3 & 100\\
\hline
14 & 0 & 9.1 & 56.1 & 96.2 & 100\\
\hline
15 & 0 & 14.5 & 69.6 & 99.0 & 100\\
\hline
16 & 0 & 12.9 & 62.6 & 96.2 & 100\\
\hline
17 & 0 & 11.6 & 58.7 & 98.3 & 100\\
\hline
18 & 0 & 12.4 & 65.0 & 95.7 & 100\\
\hline
19 & 0 & 11.0 & 61.2 & 97.7 & 100\\
\hline
20 & 0 & 19.5 & 67.4 & 99.1 & 100\\
\hline
21 & 0 & 15.5 & 63.2 & 97.2 & 100\\
\hline
22 & 0 & 11.8 & 59.6 & 98.7 & 100\\
\hline
23 & 0 & 11.1 & 64.7 & 96.7 & 100\\
\hline
24 & 0 & 10.1 & 61.3 & 98.1 & 100\\
\hline
25 & 0 & 18.6 & 68.5 & 99.2 & 100\\
\hline
26 & 0 & 16.9 & 64.2 & 97.8 & 100\\
\hline
27 & 0 & 15.1 & 59.7 & 98.8 & 100\\
\hline
28 & 0 & 14.1 & 65.9 & 97.4 & 100\\
\hline
29 & 0 & 12.0 & 62.8 & 98.5 & 100\\
\hline
30 & 0.5 & 17.2 & 67.2 & 99.2 & 100\\
\hline
31 & 0.2 & 15.8 & 64.0 & 98.1 & 100\\
\hline
32 & 0.1 & 14.5 & 61.0 & 98.9 &100\\
\hline
33 & 0 & 13.5 & 65.0 & 97.8 & 100\\
\hline
34 & 0 & 12.5 & 62.2 & 98.6 & 100\\
\hline
35 & 0.9 & 15.8 & 66.9 & 99.3 & 100\\
\hline
\end{tabular}

\smallskip

\caption{The share of sequences $s$ for which
$G(s)\!<\!H(m(s))$.}
\end{center}
\end{table}

\paragraph{The refinement over $H(\bar{m})$ by $G(s)$ when $G(s)\!\le\!H(\bar{m})$.}
The Table 2 shows how much the estimation $G(s)$ may be more accurate than $H(m(s))$
for different values of $\mu(s)$. For such $s$ that $G(s)\!<\!H(m(s))$, let
$$\Delta(s)\!\stackrel{\text{\tiny dn}}{=}\!(H(m(s))-G(s))/H(m(s))\times 100,$$ i.e. $\Delta(s)$ is an improvement
of $G(s)$ over $H(m(s))$ in percents. We present the mean and the maximum values of the 
improvements for $m\!=\!\overline{10,35}$ and for different values of $\mu(s)$. While the 
value of $H(m(s))$ belongs to the interval $[2.93,4.15]$ for all of the sequences $s$ 
that may be defined on $[m]$, $10\!\le m\!\le\!35$, it may be seen that, for 
large enough values of $\mu(s)$, the improvement $\Delta(s)$ can be tens of percents.

\begin{table}[h!]
\begin{center}
\begin{tabular}{|*{11}{c|}}\hline
\multirow{2}*{$m$} & \multicolumn{2}{|c|}{$\mu\in(0,0.2]$} & \multicolumn{2}{c|}{$\mu\in(0.2,0.4]$} &
\multicolumn{2}{c|}{$\mu\in(0.4,0.6]$} & \multicolumn{2}{c|}{$\mu\in(0.6,0.8]$} & 
\multicolumn{2}{c|}{$\mu\in(0.8,1]$}\\ 
\cline{2-11}
& mean$\Delta$ & \ $\max\Delta$ & mean$\Delta$ & \ $\max\Delta$ & mean$\Delta$ & \ $\max\Delta$ & mean$\Delta$ & \ $\max\Delta$ & mean$\Delta$ & \ $\max\Delta$\\
\hline
\hline
10 & 0 & 0 & 12.3 & 18.4 & 21.3 & 42.9 & 30.9 & 55.8 & 53.3 & 65.9\\
\hline
11 & 0 & 0 & 8.9 & 14.9 & 20.1 & 40.6 & 27.9 & 53.2 & 45.5 & 66.9\\
\hline
12 & 0 & 0 & 6.7 & 12.0 & 19.1 & 45.4 & 29.6 & 55.8 & 47.3 & 67.8\\
\hline
13 & 0 & 0 & 10.9 & 24.2 & 19.1 & 43.6 & 27.9 & 58.0 & 48.8 & 68.6\\
\hline
14 & 0 & 0 & 10.2 & 21.8 & 20.1 & 47.4 & 29.7 & 59.8 & 50.1 & 69.2\\
\hline
15 & 0 & 0 & 13.9 & 30.6 & 20.7 & 50.5 & 31.1 & 61.3 & 51.3 & 69.9\\
\hline
16 & 0 & 0 & 13.1 & 28.6 & 19.7 & 49.2 & 28.3 & 59.7 & 45.8 & 70.4\\
\hline
17 & 0 & 0 & 11.7 & 26.8 & 19.1 & 51.8 & 29.7 & 61.2 & 47.0 & 70.9\\
\hline
18 & 0 & 0 & 12.8 & 33.6 & 20.0 & 50.7 & 28.0 & 62.4 & 48.0 & 71.4\\
\hline
19 & 0 & 0 & 12.4 & 32.0 & 19.5 & 52.9 & 29.2 & 63.5 & 48.9 & 71.8\\
\hline
20 & 0 & 0 & 12.7 & 37.5 & 20.8 & 54.9 & 30.4 & 64.5 & 49.8 & 72.2\\
\hline
21 & 0 & 0 & 12.2 & 36.0 & 19.8 & 54.0 & 28.2 & 63.4 & 45.5 & 72.6\\
\hline
22 & 0 & 0 & 12.3 & 34.8 & 19.2 & 55.7 & 29.2 & 64.3 & 46.4 & 72.9\\
\hline
23 & 0 & 0 & 13.5 & 39.3 & 19.9 & 54.9 & 27.6 & 65.2 & 47.1 & 73.2\\
\hline
24 & 0 & 0 & 13.2 & 38.1 & 19.4 & 56.4 & 28.6 & 65.9 & 47.8 & 73.5\\
\hline
25 & 0 & 0 & 12.9 & 42.0 & 19.8 & 57.8 & 29.6 & 66.6 & 48.5 & 73.8\\
\hline
26 & 0 & 0 & 12.3 & 40.9 & 19.2 & 57.1 & 27.7 & 65.8 & 45.0 & 74.1\\
\hline
27 & 0 & 0 & 12.0 & 39.9 & 19.0 & 58.4 & 28.6 & 66.5 & 45.7 & 74.3\\
\hline
28 & 0 & 0 & 12.3 & 43.2 & 19.2 & 57.7 & 27.1 & 67.1 & 46.3 & 74.5\\
\hline
29 & 0 & 0 & 12.2 & 42.3 & 18.7 & 58.9 & 27.9 & 67.7 & 46.9 & 74.8\\
\hline
30 & 2.7 & 6.8 & 13.1 & 45.2 & 19.4 & 59.9 & 28.7 & 68.2 & 47.4 & 75.0\\
\hline
31 & 2.2 & 5.5 & 12.7 & 44.4 & 18.8 & 59.4 & 27.2 & 67.5 & 44.5 & 75.2\\
\hline
32 & 1.7 & 4.3 & 12.4 & 43.6 & 18.4 & 60.4 & 28.0 & 68.1 & 45.0 & 75.4\\
\hline
33 & 1.3 & 3.1 & 12.4 & 46.2 & 18.9 & 59.9 & 26.6 & 68.6 & 45.5 & 75.5\\
\hline
34 & 0.9 & 2.0 & 12.2 & 45.4 & 18.5 & 60.8 & 27.3 & 69.0 & 46.0 & 75.7\\
\hline
35 & 5.5 & 11.9 & 13.3 & 47.7 & 18.8 & 61.6 & 28.0 & 69.4 & 46.5 & 75.9\\
\hline
\end{tabular}

\smallskip

\caption{Improvement of the greedy algorithm's accuracy estimation.}
\end{center}
\end{table}

To conclude these observations, we may say that $H(\tilde{m})$ is more accurate than 
$G(s)$ when $\tilde{m}$ is rather small that is to say that the collection $S$ contains 
only sets of low cardinality. But if there are high cardinality sets in $S$ and the greedy algorithm takes them into $Gr$, the estimation $G(s)$ appears to be more accurate than $H(\tilde{m})$. For the case when $\tilde{m}\!=\!m$ and $s_1\!=\!m$, $G(s)\!=\!1$ and $H(\tilde{m})\!=\!H(m)$, i.e. 
the estimation $G(s)$ is an accurate estimation of the ratio $w(Gr)/w(Opt)$, while the value 
of $H(\tilde{m})$ approaches its maximum value that gives the worst case of the logarithmic bound 
on $w(Gr)/w(Opt)$. 

\subsection{$G(s)$ and linear programming relaxation based estimation} 

SCP may be formulated as an integer programming problem. A {\it fractional cover} is a feasible solution of linear programming relaxation of the integer program for an instance. Having the optimal fractional cover $Opt_{LP}\!\in\!\mathbb{R}^n$ for $\mathcal{P}$, since $$w(Opt_{LP})\le w(Opt),$$ we may obtain the upper bound on $w(Gr)/w(Opt)$ for the instance:
$$\frac{w(Gr)}{w(Opt)}\le\frac{w(Gr)}{w(Opt_{LP})}.\eqno(5)$$ Let $R(\mathcal{P})\!\stackrel{\text{\tiny dn}}{=}\!w(Gr)/w(Opt_{LP})$. 

The {\it integrality gap} $IG(\mathcal{P})$ is a structural property of an integer programming problem instance that chracterize accuracy of its approximation using linear programming relaxation: $$IG(\mathcal{P})\stackrel{\text{\tiny dn}}{=}\frac{w(Opt)}{w(Opt_{LP})}.$$ As a consequence of high integrality gap for an  SCP instance, the estimation (5) may be too rough.

It is not hard to obtain such instance $\mathcal{P}$ with relatively low $IG(\mathcal{P})$ that $R(\mathcal{P})\!<\!G(\mathcal{P})$, but, also, there are such instances $\mathcal{P}$ that estimation $G(\mathcal{P})$ sagnificantly improves over linear programming relaxation based estimation $R(\mathcal{P})$. 

Consider the following instance from \cite{Vazirani}. Let $m\!=\!2^k\!-\!1$, $k\!\in\!\mathbb{Z}_+$, $U\!=\!\{e_1,\ldots,e_m\}$, where elements $e_i$ of $U$ are $k$-dimensional vectors over $\mathbb{GF}_2$. Each of the vectors $e_i\!=\!(e_{i1},\dots,e_{ik})$ is a binary representation of the number $i$, i.e. $i\!=\!\sum_{j=1}^k2^{j-1}e_{ij}$. Let $S_i\!=\!\{e_j:e_i\cdot e_j\!=\!1\}$ where $e_i\cdot e_j$ denote the inner product of the vectors. Thus, for the instance, $S\!=\!\{S_1,\ldots,S_m\}$. $w(S_i)\!=\!1$ for $i\!=\!\overline{1,m}$. 

For such instance $\mathcal{P}$, we have a fractional cover $$Cvr_{LP}=\biggl (\frac{2}{m+1},\ldots,\frac{2}{m+1}\biggr ),$$ $Cvr_{LP}\!\in\!\mathbb{R}^m$. $w(Cvr_{LP})=2m/(m+1)\!<\!w(Opt)$, $w(Gr)\!=\!k$. $IG(\mathcal{P})\!>\!0.5\log_2m$. In the Table 3, there are shown lower bounds on values of $IG(\mathcal{P})$ and $R(\mathcal{P})$ and the values of $G(\mathcal{P})$ for $k\!=\!\overline{5,10}$ (in brackets, after the value of $k$, we show the dimension $m$ of the instance). It shows that the value of $G(\mathcal{P})$ basically stays the same for increasing values of $k$, while $R(\mathcal{P})$ is greater than $G(\mathcal{P})$ and the difference between them tends to grow as $k$ grows.

\begin{table}[h!]
\begin{center}
\begin{tabular}{|c||c|c|c|c|c|c|}
\hline
$k\ (m)$ &\  5\ (31)\  &\  6\ (63)\  &\  7\ (127)\  &\  8\ (255)\  &\  9\ (511)\  &\  10\ (1023)\ \\	
\hline
\hline
$IG(\mathcal{P})\!>$ & $\!2.48$ & $\!2.99$ & $\!3.49$ & $\!4.00$ & $\!4.50$ & $\!5.00$\\
\hline
$R(\mathcal{P})\!>$ & 2.58 & 3.05 & 3.53 & 4.02 & 4.51 & 5.01\\
\hline
$G(\mathcal{P})$ & 1.29 & 1.30 & 1.30 & 1.30 & 1.30 & 1.30\\
\hline
\end{tabular}

\smallskip

\caption{$IG(\mathcal{P})$, $R(\mathcal{P})$ and $G(\mathcal{P})$ for the instances from \cite{Vazirani}.}
\end{center}
\end{table}

\paragraph{Computational costs of estimating $w(Gr)/w(Opt)$.} Note that, despite the polynomial complexity of obtaining $w(Opt_{LP})$, it is hard to compute $w(Opt_{LP})$ in reasonable time for real-world problems with $m\!\ge\!4500$ \cite{UmetaniYagiura}. Subsequently, dealing with high dimensional instances, it is computationally hard to compute $R(\mathcal{P})$ or, using branch and bound methods, it is hard to obtain the lower bounds on optimal cover's weights for subproblems' instances. Whilst the computational complexity of obtaining $G(s)$ is $O(m^2n)$.

\section*{Conclusions} 

Considering the set cover problem, we estimate the accuracy of the greedy algorithm 
for a given input. We show that the accuracy may be estimated more precisely than approximation guarantee suggests if we take into account the algorithm operating on the instance. We compare the presented estimation to another estimations of this type. We give such examples of the set cover problem instances that the presented estimation sagnificantly improves over linear programming relaxation based estimation.

%
%
%

%
\clearpage

\begin{thebibliography}{9}
{\small

\bibitem{Garey} Garey, M., Johnson, D. Computers and intractability: a guide to the theory of NP-completeness. W.H. Freeman \& Co. New York, NY, USA. 1990. 

\bibitem{Dinur} Dinur, I., Steurer, D. {\it Analytical approach to parallel repetition} // STOC '14: Proceedings of the forty-sixth annual ACM symposium on Theory of computing, ACM. 2013.\ ---  
pp. 624–633.

\bibitem{RazSafra} Raz, R., Safra, S. {\it A sub-constant error-probability low-degree test, and sub-constant error-probability PCP characterization of NP} //	Proceeding
STOC '97 Proceedings of the twenty-ninth annual ACM symposium on Theory of computing. 1997.\ --- pp. 475--484. 

\bibitem{Feige}
Feige, U. {\it A threshhold of $\ln n$ for approximating set cover} // J. ACM 45 (1998). No. 4. --- pp. 634--652.

\bibitem{Chvatal}
Chvatal, V. {\it  A greedy heuristic for the set-covering problem} // Mathematics of operation research. 1979.\ V.\ 4,\ No. 3.\ --- pp. 233--235.

\bibitem{Slavik}
Slavik, P. {\it A tight analysis of the greedy algorithm for set cover} // Proceedings of the twenty-eighth annual ACM symposium on Theory of computing. 1996.\ --- pp. 435--441. 

\bibitem{Vazirani} Vazirani, V. {\it Approximation Algorithms}. Springer-Verlag, ISBN 3-540-65367-8.

\bibitem{UmetaniYagiura} Umetani, S., Yagiura, M. {\it Relaxation heuristics for the set covering problem} // Journal of the Operation Research Society of Japan. 2007. V.\ 50, No. 4. \ --- pp. 350-375.

}

\end{thebibliography}
\end{document}